# H(650) → W$^+$W$^-$/ZZ predicts H$^{++}$ → W$^+$W$^+$ and H$^+$ → ZW$^+$, as indicated by LHC data

Alain Le Yaouanc[1], François Richard[2]

Université Paris-Saclay, CNRS/IN2P3, IJCLab, 91405 Orsay, France

March 2024

**Abstract**

*Several indications for neutral scalars are observed at the LHC. One of them, a broad resonance peaked at about 650 GeV, called **H(650)**, was first observed by an outsider combining published histograms from ATLAS and CMS on ZZ →4ℓ searches, and this combination shows a local significance close to 4 s.d. Since then, CMS has reported two other indications at the same mass, with similar local significances: H →WW →ℓvℓv and H(650)→bbh(125) where mbb~90 GeV and h(125) →γγ. ATLAS has completed its analysis of ZZ→4ℓ from which we infer an indication for H(650) with 3.5 s.d. significance in the cut based analysis. Assuming that the mass is already known from the former set, and combining these three results, one gets a **global statistical significance above 6 s.d.** H(650) has a coupling to WW similar to h(125) and therefore we argue that a sum rule (SR) required by **unitarity** for W+W-->W+W- implies that there should be a compensating effect from a doubly charged scalar H++, with a large coupling to W+W+ which can calculated. We therefore predict that this mode should become visible through the vector boson fusion process W+W+->H++, naturally provided at LHC. A recent indication for **H++(450)->W+W+** from ATLAS confirms this prediction and allows a model independent determination of BR(H++ ->W+W+)~10%, surprisingly low and implying the occurrence of additional decay modes H'+W+ and H'+H'+ with one or several light H'+ with masses below mH++ - mW or MH++/2, that is **mH'+ < 370 GeV or 225 GeV**. A similar analysis is performed for **H+(375)->ZW+**, also indicated by ATLAS. Both channels suggest a scalar field content like the **Georgi Machacek** (GM) model with **two triplets**, at variance with the BSM models usually considered in the Higgs sector. Implications on **precision measurements** are presented, followed by a **complete extraction of the GM parameters.** An alternate interpretation of the 650 GeV resonance as a KK **tensor** is also discussed.*

---

1 Alain Le Yaouanc <alain.le-yaouanc@ijclab.in2p3.fr>
2 Francois Richard <francois.richard@ijclab.in2p3.fr>

# I. Introduction

At the recent conference LHCP 2023 in Belgrade, the ATLAS collaboration [1] has released an analysis for the topology W+W+ + two jets, which amounts to select the fusion mechanism, VBF, pp->W+W+jj**->ℓ+ℓ+νvjj**, where two valence u quarks present in the proton emit a W+ plus a quark d, as shown diagrammatically below.

The same mechanism happens, less frequently, for W-W-, with valence d quarks.

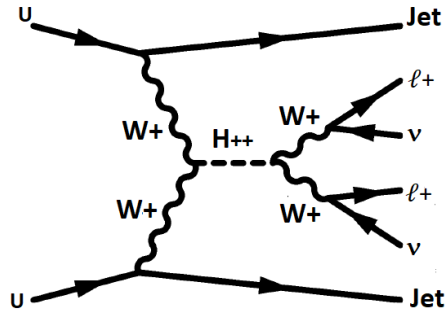

To select this reaction, the spectator quarks, dd or uu, are required to appear in the detector, while the same sign W are detected by **like sign leptons**. The WW invariant mass can be reconstructed with reasonable resolution using the standard transverse mass technique.

This analysis allows to search for doubly charged scalars appearing in **BSM models** which go beyond the usual doublet-singlet framework assumed for searching extra scalars.

Similar techniques were used in CMS to detect the W+W- final state, resulting in some evidence for a neutral scalar with a mass of 650 GeV: **pp->H(650)jj->W+W-jj->ℓ+ℓ-νvjj**. This analysis is also available in ATLAS [6] but with a reduced sensitivity given that this experiment ignores eeνv and μμνv final states.

So far, evidence for extra scalars were presented in various papers and [2] gives a summary of these effects. Recently an interpretation of these indications was presented at the ECFA workshop on this topic [3].

In [2], the significance of the various effects was crudely estimated by simply multiplying the p-values of the various channels, which is known to be an incorrect method.

In the present paper, two rigorous approaches [4,5] were used to combine the three indications for H(650), consistently reaching the same **global significance of 6.1 s.d**. In appendix A1, details are given about these calculations and the history of the evolution of the significance of H(650).

Admittedly, ATLAS shows an indication for H(650)->ℓ+ℓ-ℓ'+ℓ'- in a cut based analysis but does not confirm the effect in the ZZ->ℓ+ℓ-νv channel. Until CMS delivers its verdict on H(650)->ZZ, the initial indication obtained externally [7] by combining the histograms published by ATLAS and CMS, plays a pivotal role in two alternate ways. Either by delimiting the mass region to analyse, which avoids any degradation by the well-known "look elsewhere" effect LE, of by including this result itself, together with the published ZZ, WW and bbγγ channels and applying an LE correction. The later leads (A1) to a **global significance of 6.5 s.d.**

Recently [43], ATLAS has shown evidence for 'A(650)->H(450)Z->tt ℓ+ℓ-'. This interpretation solely relies on the 2HD expectation that mA>mH. Recall that A(420)->tt has been identified in [46] as a CP odd state. Given the poor tt mass resolution, one can expect that the two A masses are compatible.

One can therefore interpret this indication as coming from **H(650)->A(420)Z->tt ℓ+ℓ-**. The evidence is at the 2.85 s.d. level and, given that A has already been observed into top pairs and into A->H(320)Z [47], one can ignore the LE weakening.

Clearly H(650) remains the candidate with the highest significance after h(125) and RUN3 should allow to reach a firm and final conclusion about this scalar. The purpose of the present paper is to show that this particle, together with h(125), is the driving input which predicts the existence of a doubly charged scalar **H++ needed to satisfy perturbative unitarity**.

In the present work, the popular h(95) candidate is also taken into account, in spite of two short-comings: It barely reaches our 4 s.d. criterion for the global significance and official results from ATLAS, CMS or LEP2 do not provide the coupling to WW which is needed for the sum rule. We will therefore use the non-official result from [8], which combines CMS and ATLAS results for h(95)->W+W-.

H(320), seen in **A(420)->H(320)Z** [46], is uncertain for the present evaluation of SR, since its WW couplings have not been measured. As already presented at ECFA [3], our group is currently developing a method which allows to recover the missing WW/ZZ/tt couplings for H(320).

The following table summarizes the relevant indications for neutral and charged scalars. While no single channel reaches the critical global significance of 5 s.d. which, in particle physics, is required to claim a discovery, H(650) is passing this criterion if one statistically combines the four channels which have been observed so far. The details of our assumptions to reach this result are explained in A1.

In section III, we will show that, by combining h(125) to h(95) and H(650), one can predict the occurrence of a scalar with double charge as explained in [9] and [10]. This is based on a **sum rule** (SR) relating neutral scalar couplings to W+W- and the coupling of a putative doubly charged scalar H++ to W+W+, avoiding unitarity violations in the reaction W+W- -> W+W-. One can intuitively understand this as a compensating u-channel exchange balancing the s channel exchange from neutral scalars.

| Scalar | Channels | References | # s.d. glob. |
|---|---|---|---|
| H650 | WW/ZZ ggF VBF h95h125 | 1806.04529<br>2009.14791<br>2103.01918<br>CMS-PAS-HIG-20-016<br>2310.01643 | 6.1 |
| h95 | $\gamma\gamma$ $\tau\tau$ bb (LEP2) | 0306033<br>1811.08159<br>1803.06553<br>CMS-PAS-HIG-20-002<br>ATLAS-CONF-2023-035 | 3.9 |
| H++450 | W+W+ | 2132.00420<br>2104.04762 | 2.6 |
| H+375 | ZW | 2207.03925<br>2104.04762 | 2.7 |
| H++ & H+ | | | 4 |

In the next section we will list the experimental inputs needed to infer the couplings of H++ to W+W+, providing the partial width $\Gamma_{H++ \to W+W+}$ and the VBF cross section σ++.

From the experimental result giving σ++BR(H++->W+W+), one can obtain the branching ratio BR and therefore the total width of this resonance $\Gamma_{tot} = \Gamma_{H++ \to W+W+}/BR$. This BR is significantly below unity, implying additional processes of the type H3+W+ and H3+H3+, where H3+ is a charged scalar, belonging to the I=1 multiplet of GM and lighter than mH++ - mW+ or mH++/2.

A similar reasoning occurs for H+->ZW+ using a sum rule relating the H+->ZW+ coupling to the product of the couplings to ZZ and WW of the neutral scalars.

## II.     Indications for $H^{++}(450)$ and $H^{+}(375)$

These two resonances appear in models like the **Georgi Machacek model GM**. The couplings to W+W+ and ZW are simply proportional to the common vacuum expectation of the two triplets of GM that we call **u**. This quantity is trivially related to the $s_H$ parameter used by CMS to interpret these data: $s_H = 2\sqrt{2}u/v$, where v=246 GeV. This remains true in the extended version of **e-GM** that was considered [3] to interpret the properties of H(650), where we added an extra doublet to GM.

ATLAS and CMS have produced the following limits:

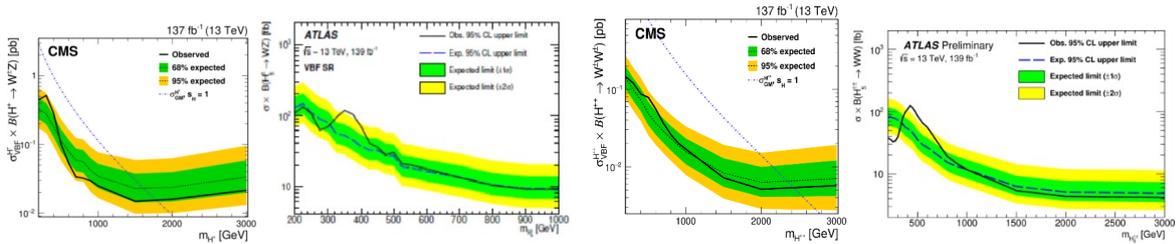

*Figure 1: Results from ATLAS and CMS on their searches for ZW and W+W+ using leptonic channels. The blue dotted curve from CMS plots correspond to the maximum expected cross sections assuming $s_H$=1, BR(WW)=1 and BR(ZW)=1.*

The first and second plot are for H+->ZW in CMS [11] and ATLAS [12]. The blue dotted curve in the CMS plot is a GM prediction which assumes $s_H$=1 and a BR(ZW)=1. These predictions scale like $(s_H)^2$BR. Both experiments show an indication near 400 GeV. ATLAS results are more recent, with a larger efficiency. They report a 2.8 sd excess around 375 GeV.

While [12] does not provide directly the measured cross section, it can be deduced from their GM interpretation of the excess as due to GM with $s_H$=0.5 and BR=0.5 (see figure 6 of [12]), which allows to reconstitute the measured **elastic VBF cross section** σ(H+->ZW+)=1000fb$(s_H)^2$BR=125±50 fb, the latter being a measurement, therefore model independent.

The third and fourth plot are for H++->W+W+ in CMS [11] and ATLAS [1]. ATLAS reports a 3.2 local excess at 450 GeV. The difference between the masses of H++ and H+ may not be significant. This excess would correspond to $s_H$=0.24 with BR=1, hence σ(H++->W+W+)=1300fb$(s_H)^2$=75±30 fb which, again, corresponds to the measured VBF cross section.

In the next section we will show how the couplings of H++ to W+W+ and H+ to ZW can be related to the **couplings** of the observed neutral scalars through two sum rules. From these couplings one can determine the total VBF t**otal cross sections** for H++ and H+. From the **elastic cross-section** measurements of ATLAS and CMS, one deduces the BR of these resonances into W+W+ and ZW.

# III. The two sum rules and the predictions for $H^{++}$ and $H^{+}$

The sum rule approach is described in [9] and [10]. It assumes a **perturbative** world where W/Z interact weakly which, so far, agrees with LHC observations. It is required by the **unitarity constraint**.

For **W+W- ->W+W-,** reference [14] gives its simplest expression which can be summarized by the following formulae:

$$g^2(4m_W^2 - 3m_Z^2 c_W^2) \stackrel{\rho \simeq 1}{\simeq} g^2 m_W^2 = \sum_k g_{W+W-H_k^0}^2 - \sum_l g_{W+W+H_l^{--}}^2$$

For **W+W- ->ZZ**, one has:

$$\frac{g^2 m_Z^4 c_W^2}{m_W^2} \stackrel{\rho \simeq 1}{\simeq} g^2 m_Z^2 = \sum_k g_{W+W-H_k^0} g_{ZZH_k^0} - \sum_l g_{W+ZH_l^-}^2$$

The first formula connects the sum of the squares of the W+W- couplings of the various neutral scalars to those of the doubly charged scalars. From LHC results we already know that the coupling of h(125) is very close to the SM value gm$_W$. This means that if one observes additional neutral scalars substantially coupled to W+W-, this calls for H++ with a coupling related to those of these scalars.

The second sum rule gives a similar message where one can predict the square of the couplings of a charged scalar to ZW in terms of the product of the couplings to WW and ZZ of the neutrals.

The following table summarizes our knowledge of the couplings of the neutral scalars normalized to the SM value gm$_W$.

All data come from LHC with the notable exception of h(95)ZZ which is derived from LEP2. The coupling of h(95) to WW comes from an unofficial combination of ATLAS and CMS data [8] and can be used for a cross-check.

For what concerns h(125) we assume that it is SM, an excellent approximation.

| Scalar | W+W-/SM ampl. | ZZ/SM ampl. |
|---|---|---|
| h(95) | 0.7±0.15    [8] | – 0.35±0.1   LEP2 |
| h(125) | ~1 | ~1 |
| H(320) | – 0.4      matrix | – 1.2  matrix |
| H(650) | – 0.9±0.15 | – 0.6±0.15 |

The ZZ/SM coupling for H(650) comes from our interpretation of the excess observed by ATLAS in VBF->H(650)->ZZ->ℓ+ℓ-ℓ'+ℓ'-.

The WW/SM coupling for H(650) comes from the CMS measurement of VBF->H(650)->WW->ℓ+νℓ-ν.

For H(320), we use the results of the **matrix method** described in section VII.

One derives:

$$g^2_{H++WW}/SM=1.3\pm0.4 \quad g^2_{H+ZW}/SM=0.8\pm0.2$$

hence the partial width of elastic modes:

$$\Gamma_{H++WW(450)}=15\pm5 \text{ GeV} \quad \Gamma_{H+ZW(375)}=12\pm4 \text{ GeV}.$$

From these and the SR, one can deduce the total cross section, the elastic BR and the total widths as given in the following table:

| Channel | $\sigma_{VBF}$ fb | $\sigma_{VBF}$ VV fb | BR(VV) % | $\Gamma$tot GeV |
|---|---|---|---|---|
| H++(450) | 830 | 75 | 9±4 | 160 |
| H+(375) | 810 | 125 | 15±8 | 80 |

Are these results **model independent ?** They assume that there is only one H++ and no other neutrals than h95, h125, H320 and H650.

Taken at face value, they indicate that the elastic channels W+W+ and ZW+ are not the dominant decay modes. This can happen in a various BSM models, among which the GM model.

## IV. A GM interpretation

From the sum rules, one is able to deduce the coupling constants $g_{H++W+W+}$ and $g_{H+ZW+}$, which can be interpreted, within the GM model in terms of $s_H$, or equivalently u the triplet vacuum expectation,.

**H+(350)** and **H++(450)** would belong to the I=2 H5 multiplet and decay into the I=1 H3 multiplet, which should contain a charged scalar H3+ and a CP odd A3 state, which should be light enough to allow dominant transitions like H5+(375)->H3+Z, A3W+ and H5++(450)->H3+W+, therefore explaining the low BR.

There is a H3+(130) candidate [20], seen from top pairs, in the bc mode by ATLAS, as discussed in section V. One expects a CP odd A3 at about the same mass. Using the measured values of the vacuum expectation u~70 GeV deduced from the SR, one can then predict:

$\Gamma_{H5++->W+W+}$=**15 GeV** $\Gamma_{H5++->H3+W++}$=**20 GeV** $\Gamma_{H5+->ZW+}$=**12 GeV** $\Gamma_{H5+->H3+Z}$ = **7.5 GeV** $\Gamma_{H5+->A3W+}$ = **6 Gev**

The following table summarizes our results in terms of the GM interpretation.

| Channel | u GeV | $s_H$ | BR(VV) % | BR(VH) % |
|---|---|---|---|---|
| H++ | 70±12 | 0.80±0.1 | 9 | 12.5 |
| H+ | 80±13 | 0.90±0.2 | 15 | 17 |

The values derived for u from the two resonances are compatible within errors. A scenario with only **one triplet** [42] is incompatible which such a large value of u which would violate the **ρ parameter** constraint.

These GM BR do not sum up to the predicted total widths, implying **large triple scalar contributions** like H5++->H3+H3+ and H5+→A3H3+. Appendix A5 shows why this explanation holds, providing the adequate BR(HH).

Again, one may worry about **model dependence**. With a **SUSY extension of GM** [49], one has 3 triplets, meaning that the SR contributions from the neutrals is shared between two H++, meaning that the coupling of H++(450) to W+W+ could be smaller. This in turn implies that **u could be smaller** and also that BR(W+W+) should be larger to reproduce the observed cross section.

Assuming a genuine GM interpretation, these results suggest a **line of action** using the VV and HV modes which would allow to simultaneously detect H5++, H5+, H3+ and A3. One can select VBF candidates tagged by the presence of Z or W decaying leptonically, associated with 2 extra jets, at least one of them being a b quark. A two-dimensional mass plot should then indicate H+(375) and H++(450) signals and reveal the presence H3+(130)->bc and its partner A3(151)->bb.

Discovering the HH modes of H++ and H+ is certainly much more challenging since the final states are dominated by 4 jet events with the presence of b/c quarks.

In addition to H5++ and H5+, one expects a neutral scalar H5 predominantly containing triplet fields but not necessarily identical to H5 since it can acquire, through mixing, doublet components.

**H(650)** is not a valid candidate since :

- it is much heavier than H++
- It is not predominantly made of triplets (see section VII)
- it gives BR(ZZ)/BR(WW)~0.2 instead of 2, as expected for a genuine H5

To interpret H(650), one needs to extend GM with an additional doublet [2], H(650) being predominantly formed by a combination of the two doublets fields.

More promising seems **H(320),** observed through the cascade A(420)->H(320)Z. Its mass seems right but the decay mode into h(125)h(125)->bbbb appears incompatible with a genuine GM which predicts decoupling between H5 and the singlets H1 and H'1 expected to constitute h(125) and h(95). This statement however ignores that a **neutral scalars are subject to mixing**, which can bring in doublet components into H(320) and also modify its triplet composition. In section VII, it will be shown that indeed the composition of H(320) differs from a genuine H5 although, contrarily to H(650), it is dominated by a triplet component. This happens as well for **h(125),** which contains triplet components without violating precision measurements, as discussed in section VII.

These mixing effects allow H(320)->h(125)h(125), h(95)h(95) and h(95)h(125). Both scalars would presumably contribute to H(320)->bbbb, given the **poor mass resolution** of this final state. The **allowed GM transition** H(320)->A3(151)A3(151) (see section VI) is even more likely to contribute to the same final state, being an allowed mode in GM.

**Mixing** effects may also affect the **charged sector**, recalling that in e-GM one expects three charged states H+. Again, these states are likely to differ from the genuine GM model. For instance, an H5+ from the I=2 multiplet could contain doublets fields which would then also allow decays into tb. Similarly, an H3+ from I=1 could contain triplet constituents allowing decays into ZW+.

In conclusion, H(320) seems a reasonable candidate to complete the I=2 multiplet with H++(450) and H+(375).

## V.     Evidence for H+(130) and A(151)

There is an indication for t->H3+(130)b with H3+->cb [20], which could contribute to H++ decays.

This prediction is inconsistent with the limit set by b->s$\gamma$ for 2HD models of type II but can be accommodated in **type I** [21] for $\tan\beta>2$.

There is a A/h(151) candidate observed into two photons, requiring additional features like transverse missing energy which can be attributed to the presence of a Z decaying into two neutrinos. It could be interpreted as the cascade H(650)→A(151)Z. In this process Z receives a large transverse momentum which easily fulfil the selection. BR(A→2γ) can still be substantial given the mass of A.

An extended e-GM model with an extra doublet, comprises three H+ and one H++. They can generate measurable deviations in the γγ **couplings** of h(95) and h(125).

The μ ratio normalized to the SM value is compatible with 1 for h125 while for h95 one measures [8]

$$\mu_{95\gamma\gamma}=0.27+0.1-0.09$$

Interpreting these results is a challenge for 2HD models and, for what concerns GM, it would require reconstructing the scalar potential in e-GM, a non-trivial task. We will come back to this issue in section VII and in appendix A5.

## VI.     Summary of findings and expectations

The following diagram summarizes the present list of significant findings [2].

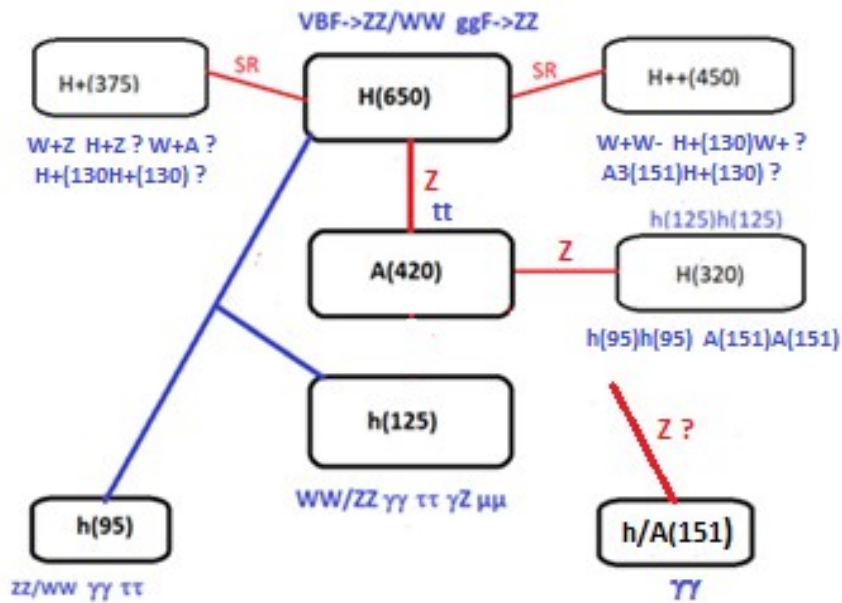

Note the recent addition [43] of the channel H(650)->AZ.

It is legitimate to ask oneself if there are searches with **negative results** which contradict some of the indications reported in this note.

Concerning the **H(650)->ZZ** channel, while ATLAS result [23] agree with the first evidence [7] with a CBA analysis using the 4-leptons, it reaches a negative limit in tension with this result by using the MVA analysis and adding events with two neutrinos. CMS has delayed its answer, which is promised to appear soon.

Concerning the H(650)->W+W- channel, as already mentioned, ATLAS does not reach the same sensitivity as CMS since it only considers the μeνν combination [6].

For what concerns H(650)->h(95)h(125)->bb$\gamma\gamma$, the recent search presented by ATLAS [24] for X->h(125)S, does not apply since it is limited to mS>200 GeV, to allow for the opening of S->ZZ/WW.

**A/h(151)**, ignored in our matrix analysis of section VII, can be interpreted, within GM, as a CP odd particle which would be the companion of H3+(130) in the I=1 multiplet. Recall that it is observed requiring missing energy which could result from H(650)->A3(151)Z, with Z->$\nu\nu$.

**A(420)** is unlikely to originate from the I=1 GM multiplet but, as for H(650), could originate from the presence of an additional doublet in e-GM.

An **additional charged Higgs** coming from the extra doublet is expected, not indicated by the data. Given that A(420) and H(650) most likely originate from the extra doublet sector, it is tempting to believe that the missing H+ has a mass in between 400 and 650 GeV, allowing decays like tb, AW+ or even ZW+ since, by mixing, H+ contains triplet constituents.

In summary one should:

- Establish beyond any doubt H(650) with the near prospect of a result of CMS on the ZZ channel
- Confirm the reaction H++->W+W+ which definitely eliminates standard explanations for new scalar resonances and try to observe H++->H3+(130)W+
- Similarly, confirm H+->ZW+ and try to observe H3+(130)Z and A3(151)W+
- Confirm the channels A(420)->tt, A(420)->H(320)Z and H(650)->A(420)Z
- Confirm the h(95) resonance
- Discover new indirect effects like an anomalous triple Higgs coupling
- Etc …

## VII. What about precision measurements?

Figure 2 shows that **B physics** constraints [27] are in tension with our solution which suggests that the value of u deduced from the SR could be too large.

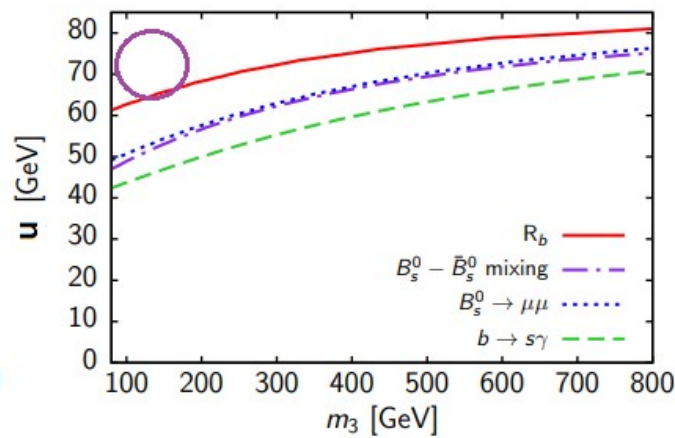

*Figure 2: This picture, taken from [27], indicates the upper limits (2 s.d. level) derived on the parameter u of the GM model from B physics PM. The ellipse shows our prediction.*

The recent measurement of **M$_W$ by CDF** has been interpreted as evidence in favour of GM, this model being able to generate a small deviation of the $\rho$ parameter – or equivalently of the **T parameter** (see for instance [25]). This effect can indeed be explained by misaligned vacuum expectation values of GM triplet constituents, with custodial symmetry breaking term in the potential, another example of extension for this model.

Alternatively, reference [26], using [27] results, suggests that within GM a significant shift of Mw can be achieved through the **S parameter**. This reference assumes that u<30 GeV, based on B physics constraints, which limits the effect. Our solution says that u~70 GeV, which increases the shift of Mw. Even with this increased value of u, the effect on the S parameter remains compatible with standard Mw measurements, being unable to accommodate the result from CDF.

In GM, the **couplings of h(125)** are usually parametrized in terms of a mixing angle $\alpha$ which characterizes the amount of the non-doublet part in h(125) and u, the common value of the vacuum expectation of the two triplet constituents. A conventional wisdom for the GM model suggests that to satisfy the PM which show **no significant deviation from the SM**, h(125) needs to be dominated by its doublet component, hence a low value of $\alpha$. This is shown in figure 3 from [28], which predicts (dark grey area) u<50 GeV and -30°<$\alpha$<10°, noting however that this plot shows an alternate solution with **u=77 GeV** and a large mixing angle $\alpha$**~60 degrees**, suggesting that h(125) may **contain** a **large triplet component.**

In our work presented at ECFA [3], an attempt was made to interpret the properties of H(650) and we came to a solution - not unique - shown in the table below, which again demonstrates that the PM can be passed with a h(125) containing triplet constituents. Indeed, this solution shows that h(125) comprises not only the two doublet components $\phi1$ and $\phi2$ but also contributions from the two triplet constituents $\chi$ and $\xi$. One also finds the vacuum expectations v1=-30 v2=102, for the two doublets and u=69.5 GeV for the triplet constituents, the later falling close to the value given by the sum rule in section III and to the solution shown in figure 3.

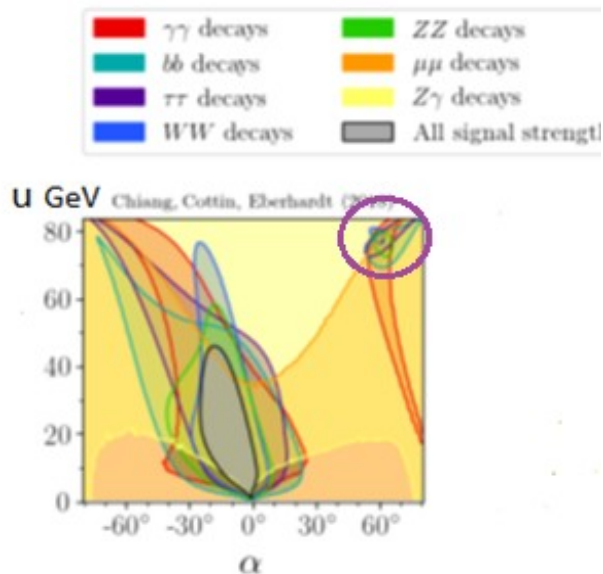

*Figure 3: Impact of various h(125) signal strengths in the u/$\alpha$ plane showing a large u/$\alpha$ solution for u=77 GeV and $\alpha$=60° (region surrounded by a magenta circle), where u is the vacuum expectation of the GM triplets, while $\alpha$ characterizes the mixing angle of the triplet constituents of h(125).*

One has v/√2=174 GeV=sqrt(v1²+v2²+4u²). A type I solution is assumed for the Yukawa couplings.

|      | 1    | 2     | 3    | 4     | Ytt/SM | ZZ/SM | WW/SM |
|------|------|-------|------|-------|--------|-------|-------|
|      | φ1   | φ2    | χ    | ξ     |        |       |       |
| H95  | 0.10 | - 0.56| 0    | 0.80  | - 0.96 | -0.34 | 0.60  |
| H125 | 0.58 | 0.58  | 0.47 | 0.33  | ~1     | ~1    | ~1    |
| H320 | 0.27 | 0.34  | -0.80| 0.35  | 0.50   | -1.20 | -0.40 |
| H650 | 0.79 | -0.50 | -0.1 | -0.34 | -0.90  | -0.60 | -0.90 |

This matrix method allows to predict (coloured squares) some yet unmeasured couplings like h(95)WW which agrees, within errors, with the unofficial result [8] used in the present analysis. For H(320), it provides the 3 couplings.

These results closely follow [2] with the exception of VBF→H650→ZZ where we assume a cross section twice as large.

Above table delivers an important message: **All neutral scalars are a substantial mixture of triplet and doublet components**, resulting in states which are not the naïve GM states.

For instance, this **matrix** tells us that h(125) is made almost equally of the two iso-doublets and of the singlet formed with triplet constituents H1'=√(1/3)ξ + √(2/3)χ :

$$h(125)=0.58(\phi_1+\phi_2)+0.58H'_1$$

This situation offers a challenge to precision measurements and one may, for instance, search for a convincing proof of the presence of H'1 in h(125).

For the **triple Higgs coupling** large deviations from the SM are still allowed. In appendix A5 we predict that λ**hhh/SM=5**, a value already testable at HL-LHC.

These effects may also become manifest in the 2γ and Zγ couplings. Recently [45] a possible deviation of h(125)->Zγ has been reported. In appendix A5 we show that there is a good agreement between our prediction and the observed value.

With **μ95γγ=0.27±0.1,** above solution requires a very significant cancellation from the charged scalar sector since, without this contribution and given the flip in sign of the h95tt coupling, one predicts μ95γγ~1. This requirement seems a priori unlikely in absence of effect for μ125γγ. Recall however that these two states have very distinct compositions. Solving this issue requires a complete knowledge of the parameters of the scalar potential, a stage one is gradually reaching (see appendix A5).

# VIII.   A T(650) solution ?

One could envisage another scenario: some of these resonance could be **tensor particles**!

This scenario naturally appears in extra dimensions where such tensors would be KK excitations of the **graviton**. If one assumes that T(320) is the first KK excitation of the graviton, one finds that T(650) can be interpreted as the **second KK excitation**, which quantitatively explains the large mass gap (see appendix A4) not easily interpreted within GM.

Given the properties of the 650 GeV resonance, it is tempting to select a Randall Sundrum type scenario where:

- T(650), peaking at the Higgs brane, naturally couples to $W_LW_L/Z_LZ_L$, giving a large VBF cross section, with a geometrical suppression of the ggF component in the so-called **bulk scenario,** where the gluon is flatly distributed in the bulk, with small overlap with the KK graviton

- T(650) couples to the radion component of the light scalars which allows to interpret the modes T(650)->h(125)h(95) and T(320)->h(125)h(125)

- One speculates that h(95) itself is primarily a **radion**

Such a scenario could therefore embed some of the candidates presented in section VI.

What are the observables consequences? Appendix A4 provides some inputs for a comprehension of this scenario.

The **angular distributions** of T(650)->ZZ/WW and T(320)->hh should not be flat. For the VBF process, one predicts [31] **d$\sigma$/dcos$\theta$~(3cos²$\theta$-1)²,** hence a word of caution when the angular anti DY background selection is applied [32].

In this scenario one would have BR(ZZ)/BR(WW)=0.5 compatible, within errors, with observation. T(650), with a large width, could be a KK graviton recurrence of **rank 2**. A rank 3 graviton would have a mass of 950 GeV, outside of the GM expected mass range.

Once the width and the mass of this resonance are fixed, [31] allows to compute the expected rates for ZZ, WW and hh for the VBF processes. These fall well below the measurements, indicating an important ggF contribution. This contribution should come from the coupling to top pairs which is ignored in [31].

What would become of the SR? The answer to this question is non-trivial and opinions vary, often claiming that for tensors the theory is non renormalizable and therefore the SR are meaningless. With the **unitarization procedure** described in [31], one can perhaps control the large s behaviour of the cross section and it becomes legitimate to assume that the Born term can describe the tensor amplitude.

# IX. Summary and conclusions

The statistical **significance of H(650)** is constantly progressing and remains the strongest evidence for BSM physics at LHC. There are however some tensions which require mutual verifications between ATLAS and CMS.

If confirmed, this discovery opens a new territory which needs to be explored before one can draw an interpretation of this result. **Phenomenology** should be at work to guide future investigations which would allow us to fully understand this sector.

Combining the most significant signals observed at LHC by ATLAS and CMS for neutral scalars, the Haber et al. **sum rules** predicts the occurrence of H++ and H+ with large couplings to W+W+ and ZW+. **H+(375)** has been indicated by both LHC experiments and there is a recent result from ATLAS for **H++(450)** comforting this hypothesis.

Combined with the SR, these **model independent observations** allow to derive BR(W+W+) ~9±4%, BR(ZW+)~15±7%. Such BR predict the presence of **additional charged scalars**, lighter than 370 GeV or 225 GeV, into which H++ (and H+) could predominantly decay. **H3+(130)** is a plausible candidate which requires confirmation. It should be accompanied by a CP odd scalar **A3(151),** already indicated into two photons.

If confirmed, these candidates provide, within GM, a quantitative explanation of the measured branching ratios, predicting BR(W+W+)=12% and BR(ZW+)=16%. This analysis also predicts an unexpected large value of the **triplet vacuum expectation u~70 GeV**.

These low branching ratios for the VV final states can be quantitatively explained, within GM, by the two other modes HV and HH provided by H+(130) and A(151), as shown in appendix A5. In this appendix we attempt to extract the complete set of parameters of the GM model. This set of parameters quantitatively interprets the BR of H5+ and H5++ into H3+A3 and H3+H3+

Generally speaking, H++(450) and H+(375) should belong to an isospin I=2 multiplet, requiring a neutral scalar with a similar mass. **H(320)** seems to be the most likely candidate. Its detection in the **4b mode**, interpreted as coming from h(125)h(125), appears to contradict the GM model unless one takes into account that neutral scalars are subject to large mixing effects, as revealed by our matrix approach which says that **h(125)** itself contains a substantial triplet constituent. Also, given its poor mass resolution, this 4b mode could also include contributions from h(95)h(95) and A3(151)A3(151).

The neutral scalar candidate **H(650)** has decay properties which require an extension of the GM model with the **addition of a second doublet**. The matrix analysis reveals that it is dominantly composed of the two doublets, a reason to discard it as a member of the I=2 multiplet of GM where particles are made of triplet constituents.

This H(650) candidate should be accompanied by a CP-odd candidate. A(420), by now observed in 3 channels, seems to be the best choice, noting the misinterpreted indication **H(650)->AZ->ttℓ+ℓ-** observed by ATLAS [43]. An additional charged Higgs is also expected, with a similar mass, not yet indicated by present data.

A major message of our **e-GM interpretation** of these indications, through the **matrix method**, shows that the neutral scalars are hybrid states constituted of a mixture of triplet and doublets components, forbidding simplistic predictions of their properties and opening a complex phenomenology for future searches.

One cannot exclude an alternate and exciting scenario, where some of these resonances are **KK recurrences of the graviton**, which could be established experimentally by measuring the **angular decay distribution** of T(650)->ZZ.

Various indications from LHC appear very promising. They have to be consolidated with more data, an **improved synergy** between the two collaborations and a **better theoretical framework**. If some of these indications are confirmed, they should strongly influence our choice for a **future e+e- collider** (see appendix A2) and the d**etector requirements** for such machines which should then be carefully optimised for final states with large jet multiplicities.

**Acknowledgements:** *We wish to express our gratitude to the ECFA conveners of the WG1-search Aleksander Zarnecki (University of Warsaw (PL)), Rebeca Gonzalez Suarez (Uppsala University (SE)), Roberto Franceschini (Rome 3 U.) for organizing a meeting on Standard and exotic Scalars at future HET factories.*

*We gratefully thank the team from Calcutta University, Anirban Kundu and Poulami Mondal, for providing innovative interpretations of the H(650) resonance.*

*We warmly thank Gilbert Moultaka from Montpellier University for various contributions to this work, in particular for what concerns the statistical evaluation of the combined LHC results and for a critical interpretation of the mass upper bounds of the scalars in GM, as derived by M. Aoki and S. Kanemura. Gilbert has also provided a deep exploration of the "matrix method" provided by the team of Calcutta.*

*Contributions of Juergen Reuter to our understanding of the tensorial world are gratefully acknowledged with his patient and immediate answering to the questions of F.R.*

*Useful discussions with Miaoran Lu from ATLAS/IJCLab are also acknowledged, in particular for signalling H(650) seen into AZ.*

*Last but not least, F.R. is very grateful to his colleague Andreas Hocker for fruitful discussions on the ZZ searches in ATLAS.*

**References:**
*[1]Measurement and interpretation of same-sign WW boson pair production in association with two jets in pp collisions at sqrt(s)=13* TeV *with the ATLAS detector*
*ATLAS Collaboration Georges Aad (Marseille, CPPM)et al. (Dec 1, 2023)*
*e-Print: 2312.00420*
*[2]* Searches for scalars at LHC and interpretation of the findings Anirban Kundu (Calcutta U.),
Alain Le Yaouanc (IJCLab, Orsay), Poulami Mondal (Calcutta U.), François Richard (IJCLab, Orsay)
Contribution to 2022 ECFA Workshop on e+e- Higgs/EW/TOP factories
e-Print: 2211.11723
Contribution to 2022 ECFA Workshop on e+e- Higgs/EW/TOP factories
*[3] Indications for new scalars and their relevance for future e+e- machines*
*Speaker*: Francois Richard *(IJCLab Orsay)*
*Standard and exotic Scalars at future HET factories*
Friday 14 apr. 13 2023, 09:00 → 14:00 Europe/*Zurich*
*Aleksander Zarnecki (University of Warsaw (PL)), Rebeca Gonzalez Suarez (Uppsala University (SE)), Roberto Franceschini (Rome 3 U.)*
*https://indico.cern.ch/event/1253605/*
*[4]* Questions and Answers
Frederick Mosteller and R. A. Fisher
The American Statistician
Vol. 2, No. 5 (Oct. ,1948), pp. 30-31 (2 pages)
Published By: Taylor & Francis, Ltd.
https://doi.org/10.2307/2681650
https://www.jstor.org/stable/2681650
*[5] Combining statistical tests by multiplying p-values*
James Theiler (Los Alamos)
Astrophysics and Radiation Measurements Group, NIS-2, 2004

*[6] Search for heavy resonances in the decay channel $W+W-$ -> $e\nu\mu\nu$ in $pp$ Collisions at √s = 13TeV using 139 fb-1 of data with the ATLAS detector*
ATLAS-CONF-2022-066 29th November 2022

*[7] Evidence of the true Higgs boson HTHT at the LHC Run 2*
*Paolo Cea (INFN, Bari) (Jun 10, 2018)*
*Published in: Mod.Phys.Lett.A 34 (2019) 18, 1950137*
*e-Print: 1806.04529*

*[8] Searching for Low-Mass Resonances Decaying into WW Bosons*
*Guglielmo Coloretti (Zurich U. and PSI, Villigen), Andreas Crivellin (Zurich U. and PSI, Villigen), Srimoy Bhattacharya(U. Witwatersrand, Johannesburg, Sch. Phys.), Bruce Mellado (U. Witwatersrand, Johannesburg, Sch. Phys. and iThemba LABS) (Feb 14, 2023)*
*e-Print: 2302.07276*

*[9] Sum rules for Higgs bosons*, J.F. Gunion, H.E. Haber, and J. Wudka, Phys. Rev. D43, 904 (1991).

*[10] What if the Higgs couplings to W and Z bosons are larger than in the Standard Model?*
*Adam Falkowski(Orsay, LPT), Slava Rychkov(UPMC, Paris (main) and Ecole Normale Superieure), Alfredo Urbano (Ecole Normale Superieure) (Feb, 2012)*
Published in: JHEP 04 (2012) 073
e-Print: *1202.1532*

*[11] Search for charged Higgs bosons produced in vector boson fusion processes and decaying into vector boson pairs in proton–proton collisions at √s=13 TeVs*
CMS Collaboration *Albert M Sirunyan (Yerevan Phys. Inst.)* et al. (Apr 10, 2021)
Published in: Eur.Phys.J.C 81 (2021) 8, 723
e-Print: *2104.04762*

*[12] Search for resonant WZ→ℓνℓ'ℓ' production in proton−−proton collisions at √s=13 TeV with the ATLAS detector*
*ATLAS* Collaboration (Jul 8, 2022)
e-Print: *2207.03925*

*[13] Custodial symmetry, the Georgi-Machacek model, and other scalar extensions*
A. Kundu, P. Mondal, and P.B. Pal
Published in: Phys. Rev. D105, 115026 (2022)
arXiv: 2111.14195.

[14] Higgs Sector Motivations for an e-minus e-minus Linear Collider
 J.F. Gunion
arXiv: hep-ph/9309226

*[15] Doubly charged Higgs bosons*
*Howard Georgi (Harvard U.), Marie Machacek (Northeastern U.) (Jun, 1985)*
Published in: Nucl.Phys.B 262 (1985) 463-477

*[16] The decoupling limit in the Georgi-Machacek model*
*Katy Hartling (Ottawa Carleton Inst. Phys.), Kunal Kumar (Ottawa Carleton Inst. Phys.), Heather E. Logan (Ottawa Carleton Inst. Phys.) (Apr 9, 2014)*
*Published in Phys.Rev.D 90 (2014) 1, 015007*
*e-Print: 1404.2640*

*[17] Unitarity bounds in the Higgs model including triplet fields with custodial symmetry*
*Mayumi Aoki (Tokyo U., ICRR), Shinya Kanemura (Toyama U.)(Dec, 2007)*
*Published in: Phys.Rev.D 77 (2008) 9, 095009, Phys.Rev.D 89 (2014) 5, 059902 (erratum)*
*e-Print: 0712.4053*

*[18] Supersymmetric Custodial Triplets*
*Luis Cort (Barcelona, IFAE), Mateo Garcia (Barcelona, IFAE), Mariano Quiros (ICREA, Barcelona and Barcelona, IFAE) (Aug 19, 2013)*
*Phys.Rev.D 88 (2013) 7, 075010*
*e-Print: 1308.4025*

*[19] Some new observations for the Georgi-Machacek scenario with triplet Higgs scalars*
*Rituparna Ghosh (IISER, Kolkata), Biswarup Mukhopadhyaya (IISER, Kolkata) (Dec 22, 2022)*
Published in: *Phys.Rev.D* 107 (2023) 3, 035031
e-Print: 2212.11688

*[20] Search for light charged Higgs boson in t→H±b(H±→cb) decays with the ATLAS detector at LHC*
*ATLAS* Collaboration
*Anna Ivina (Weizmann Inst.)* for the collaboration. (Mar 16, 2022)
Published in : PoS EPS-HEP2021 (2022) 631
Proceedings of EPS-HEP2021, 631

[21] *Weak radiative decays of the B meson and bounds on MH± in the Two-Higgs-Doublet Model*
*Mikolaj Misiak (CERN and Warsaw U.), Matthias Steinhauser (KIT, Karlsruhe, TTP)(Feb 15, 2017)*
Published in: Eur.Phys.J.C 77 (2017) 3, 201
e-Print: 1702.04571

[22] *The 95.4 GeV di-photon excess at ATLAS and CMS*
*T. Biekötter (KIT, Karlsruhe, TP), S. Heinemeyer (Madrid, IFT), G. Weiglein (DESY* and *Hamburg U., Inst. Theor. Phys. II)*(Jun 6, 2023)
e-Print:2306.03889

[23] *Search for heavy resonances decaying into a pair of Z bosons in the ℓ+ℓ−ℓ′+ℓ′−ℓ+ℓ−ℓ′+ℓ′− and ℓ+ℓ−vv¯ℓ+ℓ−vv¯ final states using 139 fb−1fb−1 of proton–proton collisions at sqrt(s)=13 TeV with the ATLAS detector*
*ATLAS Collaboration*
*Georges Aad (Marseille, CPPM) et al. (Sep 30, 2020)*
Published in: Eur.Phys.J.C 81 (2021) 4, 332
e-Print: 2009.14791

[24] *Search for a new heavy scalar particle decaying into a Higgs boson and a new scalar singlet in final states with one or two light leptons and a pair of τ-leptons with the ATLAS detector*
*ATLAS Collaboration*
*Georges Aad (Marseille, CPPM) et al. (Jul 20, 2023)*
e-Print: 2307.11120

[25] *See for instance: Explanation of the W mass shift at CDF II in the extended Georgi-Machacek model*
*Ting-Kuo Chen (Taiwan, Natl. Taiwan U.), Cheng-Wei Chiang (Taiwan, Natl. Taiwan U.), Kei Yagyu (Osaka U.) (Apr 27, 2022)*
Published in: Phys.Rev.D 106 (2022) 5, 055035
e-Print: 2204.12898

*[26]Enhancement of the W boson mass in the Georgi-Machacek model*
*Poulami Mondal (Calcutta U.)(Apr 16, 2022)*
Published in: Phys.Lett.B 833 (2022) 137357
e-Print: 2204.07844

*[27]Indirect constraints on the Georgi-Machacek model and implications for Higgs boson couplings*
*Katy Hartling (Ottawa Carleton Inst. Phys.), Kunal Kumar (Ottawa Carleton Inst. Phys.), Heather E. Logan (Ottawa Carleton Inst. Phys.) (Oct 21, 2014)*
Published in: Phys.Rev.D 91 (2015) 1, 015013
e-Print: 1410.5538

*[28]Global fits in the Georgi-Machacek model*
*Cheng-Wei Chiang (Taiwan, Natl. Taiwan U. and Taiwan, Inst. Phys.), Giovanna Cottin (Taiwan, Natl. Taiwan U.), Otto Eberhardt (Valencia U., IFIC) (Jul 27, 2018)*
Published in: Phys.Rev.D 99 (2019) 1, 015001
e-Print: 1807.10660

*[29] Growing Excesses of New Scalars at the Electroweak Scale*
*Srimoy Bhattacharya (Witwatersrand U.), Guglielmo Coloretti (Zurich U. and PSI, Villigen), Andreas Crivellin (Zurich U. and PSI, Villigen), Salah-Eddine Dahbi(Witwatersrand U.), Yaquan Fang (Beijing, Inst. High Energy Phys. and Beijing, GUCAS) Jun 29, 2023*
*e-Print: 2306.17209*
*[30] Mounting evidence for a 95 GeV Higgs boson 23 août 2023,*
*Parallel session talk Higgs Physics Joint T09+T10 Higgs Physics + Searches for New Physics*
*Thomas Biekoetter (ITP Karlsruhe)*
*https://indico.desy.de/event/34916/contributions/147467/*
*[31] Resonances at the LHC beyond the Higgs boson: The scalar/tensor case*
*Wolfgang Kilian(Siegen U.), Thorsten Ohl(Wurzburg U.), Jürgen Reuter(DESY), Marco Sekulla(Siegen U. and KEK, Tsukuba and KIT, Karlsruhe)(Oct 30, 2015)*
*Published in: Phys.Rev.D 93 (2016) 3, 03600*
*e-Print: 1511.00022*
*[32] Measurement of the Higgs boson mass and width using the four leptons final state*
*CMS Collaboration*
*CMS-PAS-HIG-21-019*
*https://cds.cern.ch/record/2871702*
*[33] Search for high mass resonances decaying into W+W- in the dileptonic final state with 138 fb−1 of proton-proton collisions at sqrt(s)=13 TeV*
*CMS Collaboration (2022)*
*Report number: CMS-PAS-HIG-20-016*
*[34] Search for a new resonance decaying into two spin-0 bosons in a final state with two photons and two bottom quarks in proton-proton collisions at sqrt(s)=13 TeV*
*CMS Collaboration Armen Tuasyan (Yerevan Phys. Inst.)et al. (Oct 2, 2023)*
*e-Print: 2310.01643*
*[35] Phenomenology of the Randall-Sundrum Gauge Hierarchy Model*
*H. Davoudiasl (SLAC), J.L. Hewett (SLAC), T.G. Rizzo (SLAC)(Sep, 1999)*
*Published in: Phys.Rev.Lett. 84 (2000) 2080*
*e-Print: hep-ph/9909255*
*[36] Warped Gravitons at the LHC and Beyond*
*K. Agashe (Syracuse), H. Davoudiasl (Brookhaven), G. Perez (Stony Brook), A. Soni (Brookhaven)(July, 2007)*
*Published in: Phys.Rev.D76:036006,2007*
*e-print: hep-ph/0701186*
*[37] Search for Higgs boson pair production in association with a vector boson in pp collisions at sqrt(s)=13 TeV*
*ATLAS Collaboration*
*Georges Aad (Marseille, CPPM ) et al. (Oct 11, 2022)*
*Published in: Eur.Phys.J.C 83 (2023) 6, 519, Eur.Phys.J.C 83 (2023) 519*
*e-Print: 2210.05415*
*[38] Search for the standard model Higgs boson at LEP*
*LEP Higgs Working Group for Higgs boson searches and OPAL and ALEPH and DELPHI and L3 Collaborations (Jul, 2001)*
*Contribution to: 20th International Symposium on Lepton and Photon Interactions at High Energies (LP 01), EPS-HEP 2001*
*e-Print: hep-ex/0107029*
*[39] Gravity particles from Warped Extra Dimensions, predictions for LHC*
*Alexandra Carvalho (ICTP-SAIFR, Sao Paulo and Sao Paulo, IFT and Lyon, IPN)(Mar 31, 2014)*
*e-Print: 1404.0102*

*[40] Search for a light radion at HL-LHC and ILC250*
*Francois Richard (Orsay, LAL)(Dec 18, 2017)*
*e-Print: 1712.06410*
*[41] KK graviton resonance and cascade decays in warped gravity*
*Barry M. Dillon (Sussex U.), Chengcheng Han (Tokyo U., IPMU), Hyun Min Lee (Chung-Ang U.), Myeonghun Park (IBS, Daejeon, CTPU) (Jun 22, 2016)*
*Published in: Int.J.Mod.Phys.A 32 (2017) 33, 1745006*
*e-Print: 1606.07171*
*[42]Combined Explanation of LHC Multi-Lepton, Di-Photon and Top-Quark Excesses*
*Guglielmo Coloretti, Andreas Crivellin, Bruce Mellado (Dec 28, 2023)*
*e-Print: 2312.17314*
*[43] ATLAS CollaborationSearch for a CP-odd Higgs boson decaying into a heavy CP-even Higgs boson and a ZZ boson in the ℓ+ℓ−ττℓ+ℓ− and vvbb final states using 140 fb−1−1 of data collected with the ATLAS detector*
*Georges Aad (Marseille, CPPM) et al. (Nov 7, 2023)*
*e-Print: 2311.04033*
*[44] Unavoidable Higgs coupling deviations in the Z2-symmetric Georgi-Machacek model*
*Carlos Henrique de Lima (Carleton U.), Heather E. Logan (Carleton U.)(Sep 17, 2022)*
*Published in: Phys.Rev.D 106 (2022) 11, 115020*
*e-Print: 2209.08393*
*[45] Evidence for the Higgs Boson Decay to a Z Boson and a Photon at the LHC*
*ATLAS and CMS Collaborations*
*Georges Aad(Marseille, CPPM)et al. (Sep 7, 2023)*
*Published in: Phys.Rev.Lett. 132 (2024) 021803*
*e-Print: 2309.03501*
*[46] Search for heavy Higgs bosons decaying to a top quark pair in proton-proton collisions at sqrt(s)=13 TeV*
*CMS Collaboration*
*Albert M Sirunyan (Yerevan Phys. Inst.)et al. (Aug 2, 2019)*
*Published in: JHEP 04 (2020) 171, JHEP 03 (2022) 187 (erratum)*
*e-Print: 1908.01115*
*[47] ATLAS CONF Note. ATLAS-CONF-2022-043. 10th July 2022. Search for Higgs boson pair production in association with a vector boson in p p collisions at sqrt(s) = 13 TeV with the ATLAS detector*
*ATLAS Collaboration*
*[48]A 95 GeV Higgs Boson in the Georgi-Machacek Model*
*Ting-Kup Chen, Cheng-Wei Chiang, Sven Heinemeyer, Georg Weiglein (Dec 20, 2023)*
*e-Print: 2312.13239.*
*[49] The Supersymmetric Georgi-Machacek Model*
*Roberto Vega(Southern Methodist U.), Roberto Vega-Morales(CAPFE, Granada and Granada U., Theor. Phys. Astrophys.), Keping Xie(Southern Methodist U.)(Nov. 14, 2017)*
*Published in: JHEP 03 (2018) 168*
*e-Print: 1711.053329.*
*[50]Constraints on the Higgs boson self-coupling with combination of single and double Higgs boson production*
*The CMS Collaboration, Nov 26 2023.*
*CMS PAS HIG-23-006*

# APPENDICES

## A1. Statistical analysis of the H(650) candidate

Recall that this signal has been observed in 4 channels. ZZ into 4 leptons (3.5 s.d. local evidence in ATLAS), WW into 2 leptons and two neutrinos (3.8 s.d. local evidence in CMS). In CMS, a bb$\gamma\gamma$ final state, where the two photons form a h(125), shows a 3.8 s.d. evidence at 650 GeV when bb form a mass compatible with h(95). While the two CMS effects where quantified in the publication, it was left to the reader to estimate the significance for the ZZ->4ℓ channel in ATLAS [23].

More recently [43], ATLAS has shown some 2.85 s.d. evidence for H(650)->AZ->tt ℓ+ℓ-.

The story of this resonance started from [7] which used the published mass distributions from ATLAS (36.1 fb-1) and CMS (77.4 fb-1) and combined them.

Though the results deduced from the [7] and [23] data are not "official" claims from LHC collaborations, they seem to us worth considering since they originate from a straightforward interpretation of publicly available mass distributions of ZZ into four leptons, a **gold-plated low background channel** which has contributed to the discovery of h(125).

In a first approach the result deduced from the LHC data collected by [7] simply serves to **fix a mass** for further searches, avoiding the usual **look-elsewhere (LE)** penalties. Noteworthy noting that [7] concluded to a significance "well above" above 5 s.d. instead of 3.8 s.d. as derived from figure 4. In our opinion this contradiction comes from a difference in the statistical treatment. Our estimate follows the "**frequentist**" criteria which is the most conservative. This high significance claimed by [7], does not consider the LE criterion which further reduces the local evidence from 3.8 local to 2.8 s.d. global evidence.

Also, the mass found in [7] is "around 700" GeV instead of 650 GeV in our case, for which we do not see an explanation.

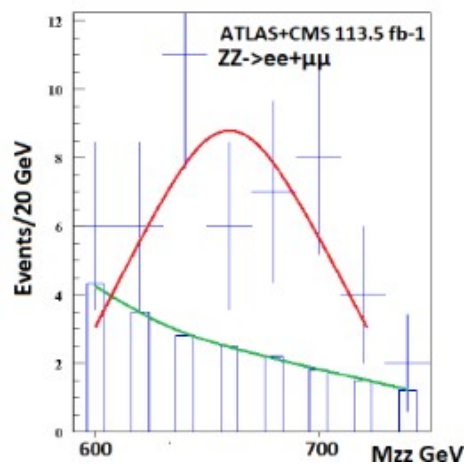

*Figure 4: ZZ->4ℓ mass distribution from [7] using a combination of ATLAS and CMS data. The blue crosses are the data. The green line describes the predicted non resonant background, and the red curve is the result of an adjustment of the data by the sum BW+background.*

The next step came from a published analysis by ATLAS for ZZ with 139 fb-1. The ZZ mass spectrum for the ggF+VBF analysis clearly shows an excess consistent in mass and width with figure 4, with a 3.5 s.d. local significance. The two samples weakly overlap since the ATLAS sample used by [7] only contains 26% of the total ATLAS sample.

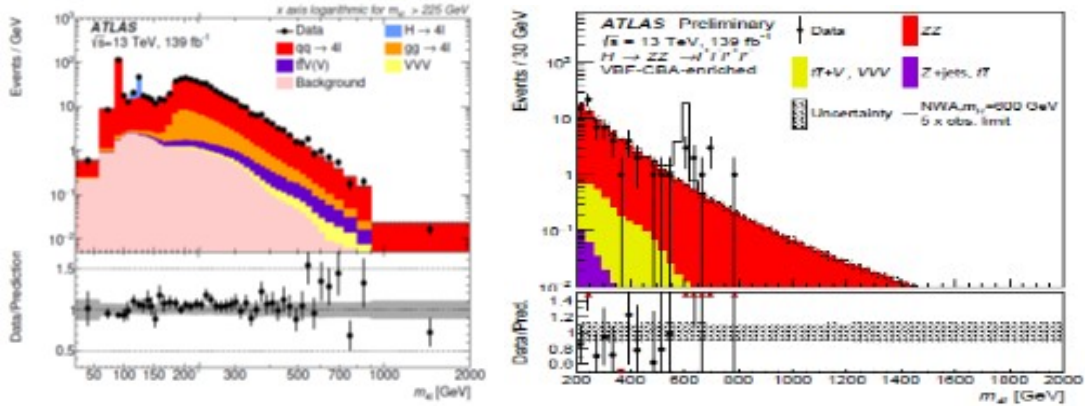

*Figure 5: ZZ->4ℓ mass distribution obtained by ATLAS [23]. The first plot assumes ggF production while the second plot selects a VBF production mechanism.*

Let us explain how the 3.5 s.d. local significance from the ATLAS is estimated from above histograms. Integrating the ggF mass distribution from 510 GeV, up to 900 GeV, one finds 228 events with a background of 179 events, hence a 3.2 s.d. local excess. The VBF part shows 9 candidates with 2 background events, hence a 2.5 s.d. local excess. Combining the two, the significance reaches 3.5 s.d., using the Fisher method.

At this point one should recall that these indications result from our interpretation of the CBA (cut based analysis) results of ATLAS and are not confirmed by the MVA (multivariate) analysis, which claims a higher purity with about the same efficiency. Adding the ℓ+ℓ-νν channels and using MVA, ATLAS [2] sets an upper limit $\sigma_{ggF}$BR(ZZ)<50 fb, in tension with the CBA result into 4 leptons $\sigma_{ggF}$BR(ZZ)=150±60 fb and $\sigma_{VBF}$BR(ZZ)=34±15 fb. CMS will hopefully soon settle this issue.

Ignoring this caveat, one is therefore left with three indications with local effect of respectively, 3.5 s.d. (ZZ), 3.8 s.d. (WW) and 3.8 s.d. (bbγγ), having a priori fixed the mass from the initial result obtained by combining ATLAS and CMS results, therefore without the need to further apply the **LE** criterion.

One can naïvely use the product of the p-values to observe the three fluctuations and derive the overall significance, which leads to a final significance of ~ 6.9 s.d.

This method is deemed incorrect and therefore we have tried two alternate methods. The Fisher method [4], which uses the combination:

$$X^2_{2n} = -2 \sum_{1}^{n} \log p_i$$

where $p_i$ is the p-value for the $i^{th}$ hypothesis test. The variable x=$X^2_{2n}$ follows the $\chi^2$ distribution:

$$x^{n/2-1} e^{-x/2} / 2^{n/2} \Gamma(n/2)$$

By integrating this quantity between the measured value $X^2_{2n}$ to infinity, one obtains the significance probability which is then translated into a Gaussian probability.

For H(650), n=3, logp1=-7.67 and logp2=logp3=-8.84, $X^2_4$=50.69 x=7.12 one finds:

$$P(X^2_4 > x^2) = 0.5\exp(-u)(2+2u+u^2) \text{ with } u=x^2/2$$

which gives **P=6.7 10⁻¹⁰** probability, that is **6.1 s.d.** global significance.

Using the method described in [5], one reaches the same conclusions.

| Steps | Mode | Origin | Local sd | Remark | Global sd |
|---|---|---|---|---|---|
| 0 | ZZ->4ℓ | ATLAS+CMS [7] | 3.8 | ATLAS+CMS 113.5 fb-1 Defines mass & width | 2.8 |
| 1 | ZZ->4ℓ | From ATLAS [23] | 3.5 | From histogram | 3.5 |
| 2 | WW->ℓνℓν | From CMS [33] | 3.8 | Official statement | 5 |
| 3 | h(95)h(125)->bbγγ | From CMS [34] | 3.8 | Official statement | 6.1 |

The historical progress of significance reached at various steps, is shown in above table. Each of these signals has its pros and cons:

- The H(650)->ZZ-> ℓ+ℓ-ℓ'+ℓ'- indication is not commented in [23] and appears to be in conflict with ℓ+ℓ-νν

- For H(650)→W+W-→ℓνℓν, CMS concludes that it is dominated by VBF, with $\sigma_{VBF}BR=160$ fb, while our estimate from ZZ suggests ggF~5VBF, with large uncertainties. This interpretation from CMS has to be taken with a 'grain of salt' given that for ggF the background is more than ten times larger than for VBF, which forbids a meaningful evaluation

- To accommodate the ggF contribution to ZZ, one requires a width $\Gamma_{tt}$~15 GeV, which would give BR(tt)~15%, hence a cross section below 100 fb, probably unobservable given the QCD background

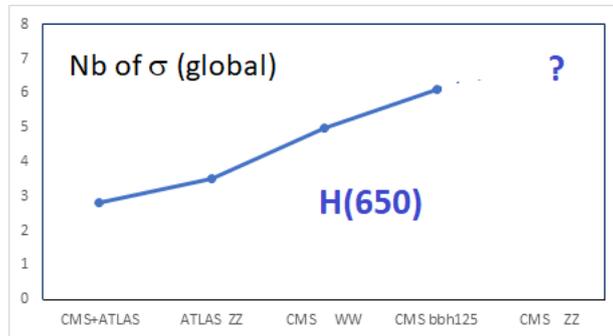

Above graphic shows the evolution with time of the evidence for H(650) starting from the ZZ evidence obtained by [7] in ZZ, noting that the verdict for ZZ from CMS is still missing.

The statistical evidence for this new resonance, above 6 s.d. global, is overwhelming. Admittedly, other interpretations of these observations are possible, but we think that the proposed picture correctly follows the historical development of our findings, which took place without an a priori knowledge of the next occurrences.

In a second approach, one fully uses the initial indication deduced from [7] for H(650) and combines it with the 3 other indications, applying a **LE** criterion. This criterion is applied to the first ZZ search reducing its significance.

To estimate the LE factor, one takes into account that the experimental data stop at 1 TeV. One defines the number of bins explored considering that the expected resonance becomes narrower at low masses. These criteria lead us to a LE factor of ~30.

Using again the Fisher method, one finds **6.5 s.d.** global significance.

The table allows to see the progress in significance reached at the various steps of this analysis.

| Steps | Mode | Origin | Local sd | Remark | Global sd |
|---|---|---|---|---|---|
| 0 | ZZ->4ℓ | ATLAS+CMS [7] | 3.8 | LE=30 | 2.8 |
| 1 | ZZ->4ℓ | From ATLAS [23] | 3.5 | From histogram | 3.5 |
| 2 | WW->ℓνℓν | From CMS [33] | 3.8 | Official statement | 5 |
| 3 | h(95)h(125)->bbγγ | From CMS [34] | 3.8 | Official statement | 6.5 |

In conclusion, it seems that both methods lead to the conclusion that the evidence for H(650) appears very strong, with **a global significance above 6 s.d**.

Admittedly, the first evidence in ZZ plays an important part in this conclusion. When CMS will release its results from RUN2, one should drop the result deduced from [7] and reach a final conclusion for RUN2 only based on the results published by the two collaborations.

## Other channels

**h(95)** is indicated in 3 channels: hZ from LEP2, γγ and ττ from CMS giving P=$10^{-5}$. This particle is searched for masses between 60 and 110 GeV in bins of ~10 GeV, leading to LE=5 and therefore has a 3**.9 s.d. global significance**. The mass domain for the two photons search starts at 60 GeV and the mass domain covered by LEP2 stops at 110 GeV. Note that the evidence for **H(650)->h(125)h(95)** from CMS is not taken into account in our estimate of the significance. We are aware that LEP2 results show an **unlikely distribution in energy** [38], most of the effect being seen at 189 GeV centre of mass energy, while one would expect a strong threshold effect favouring larger energies.

For **H(320)**, indicated in the cascade A(420)→H(320)Z, the matrix solution predicts $\Gamma_{ZZ}$~5 GeV, meaning that this signal should become visible in four leptons. To explain its absence requires $\Gamma$hh>30 GeV. For **H+(375)**, there is a 2.8 s.d. indication in ATLAS and 2 s.d. in CMS, which gives 2.7 s.d. global significance, including a LE factor of 8. This small factor reflects the poor mass resolution of this type of search and the absence of data at large masses.

For **H++(450)**, ATLAS reports 3.2 s.d. in local, which gives 2.6 s.d. global significance, including the same LE factor. This value agrees with the 2.5 s.d. global significance given by ATLAS. Combining the two measurement one reaches a **4 s.d. global significance**, including the LE effect

# A2. Expected cross-sections in e+e-

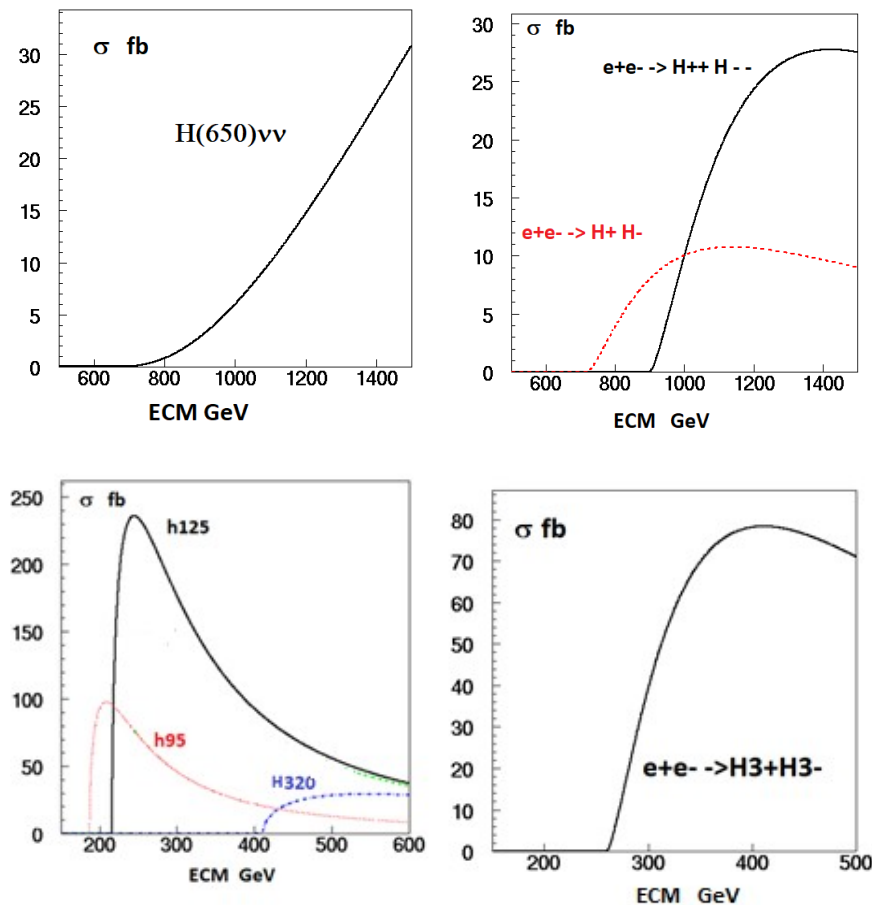

*Figure 6: Top row, on the left, e+e- cross section for producing H(650) through VBF. On the right, predicted e+e- cross-sections for H+(375) H-(375) and H++(450) H - - (450). Bottom row, predicted cross sections for the lightest scalars.*

Figure 6 shows that a circular Higgs factory operating up to the top threshold should be able to produce several of the BSM scalars. The fate of **h/A(151)** remains uncertain until one observes this particle into **WW/ZZ**, allowing e+e→h(151)Z. If this particle turns out to be **A3(151),** companion of H3+(130), it could be produced together with h(125) or h(95), therefore accessible to a circular collider.

A LC reaching **1 TeV** with high luminosity is needed to produce and accurately measure H(650), H++(450) and H+(375). H(650) is mainly produced through VBF, reaching a cross section of 10 fb at 1 TeV [2]. This figure can be doubled using an electron beam with left-handed polarisation. ILC at 1 TeV could collect 8000 fb-1 and, with polarisation, produce **$10^5$ H(650) events**.

An **e-e- collider** one could produce H- - through VBF with polarized beams ~100 fb at 1 TeV.

Final state will result into **complex topologies** requiring **an ideal detector** will full solid angle coverage, measuring and identifying a large number of jets. For instance, the cascade H(650)->A(420)Z->ttZ gives **8 jets**. It has a cross section of about ~1 fb, while the SM process ttZ has a cross section of ~5 fb at 1 TeV. This SM process could serve as a reference for optimising the properties of a future detectors at 1 TeV.

# A3. Upper mass bounds in the GM model

The appearance of a scalar resonance at 650 GeV may seem in tension with the genuine GM model. Based on unitarity constraints, [17] predicts that all GM scalars should be below 700 GeV, which is pretty close to the mass of H(650) and therefore raises some concerns.

In figure 7 from [17], one observes that mH5 is expected to be below 650 GeV. If one also assumes that mH3+=130 GeV, this mass is clearly inconsistent.

Note however that the mass bounds from [17] rely on a simplified scalar potential where the M1 and M2 terms are absent, contrary to the more general case:

$$V(\Phi, X) = \frac{\mu_2^2}{2}\mathrm{Tr}(\Phi^\dagger\Phi) + \frac{\mu_3^2}{2}\mathrm{Tr}(X^\dagger X) + \lambda_1[\mathrm{Tr}(\Phi^\dagger\Phi)]^2 + \lambda_2\mathrm{Tr}(\Phi^\dagger\Phi)\mathrm{Tr}(X^\dagger X)$$
$$+ \lambda_3\mathrm{Tr}(X^\dagger X X^\dagger X) + \lambda_4[\mathrm{Tr}(X^\dagger X)]^2 - \lambda_5\mathrm{Tr}(\Phi^\dagger\tau^a\Phi\tau^b)\mathrm{Tr}(X^\dagger t^a X t^b)$$
$$- M_1\mathrm{Tr}(\Phi^\dagger\tau^a\Phi\tau^b)(UXU^\dagger)_{ab} - M_2\mathrm{Tr}(X^\dagger t^a X t^b)(UXU^\dagger)_{ab}.$$

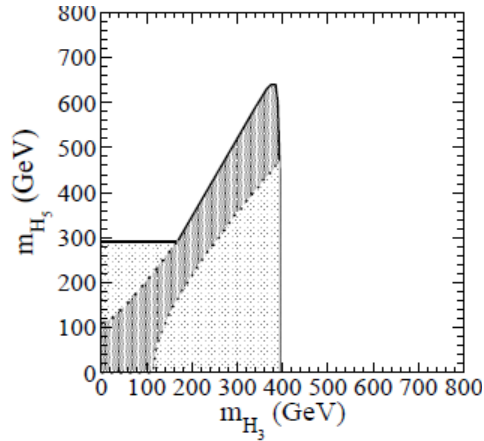

Figure 7: Allowed regions of the Higgs bosons in the mH3/mH5 plane from [17]. The light shadowed regions are excluded by the Zbb constraint.

If one introduces these terms, the unitarity limits can be widely extended as noted in [16]. Reference [44] justifies the elimination of these terms by the introduction of a Z2 symmetry which takes care of a fine-tuning effect in the determination of the Majorana mass of the neutrino but acknowledges that this requirement is by no means mandatory.

In appendix A5 we show that to interpret the properties of H5+(375) and H++(400) one has to assume **large values of M1 and M2** meaning that one cannot assume M1=M2=0, which relaxes the mass limits from figure 7.

# A4. Few inputs for a tensor scenario

Recently, CMS has produced an analysis for the reaction ZZ into 4 leptons and its interpretation in terms of the h(125) scalar boson. To improve on the signal/background ratio, this analysis has used discriminating observables against the DY background [32], with the result shown in figure 8.

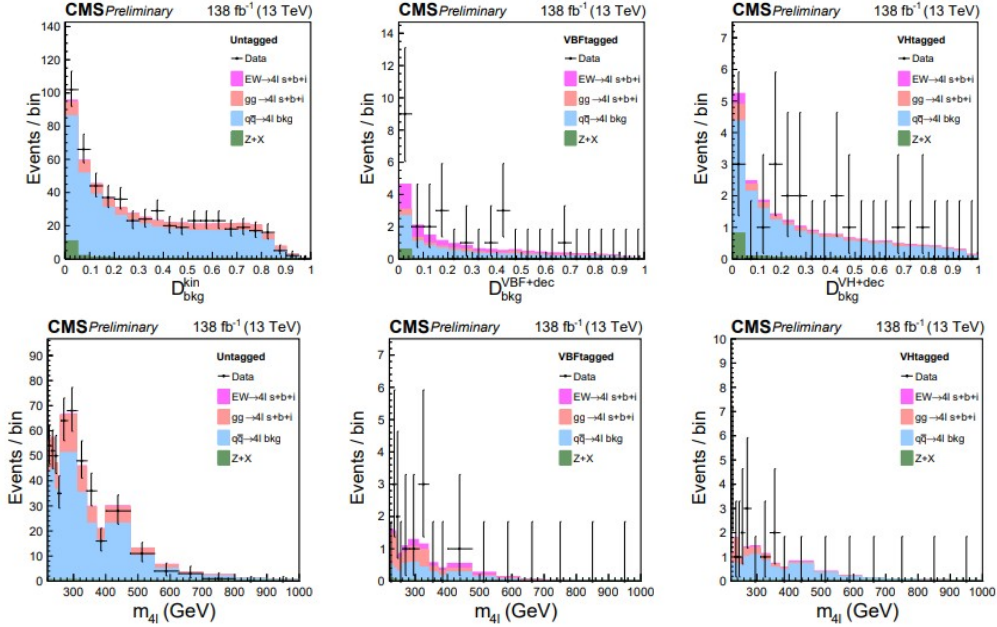

*Figure 8. . First row: distribution of the discriminant variables. To enhance a scalar signal over background contributions, a discriminant selection, D>0.6, has been applied to the three categories of the first row, meaning that these distributions only contain a reduced fraction of the total sample, therefore with reduced statistical significance. Second row: m4ℓ mass distributions observed after this selection for untagged, VBF tagged and VH tagged events, which show no excess at 650 GeV.*

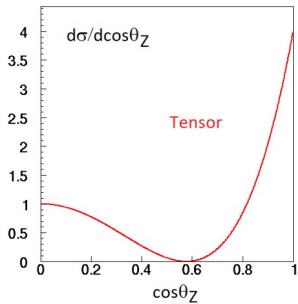

*Figure 9: Angular distribution for a tensor decaying into ZZ (ultra-relativistic approximation).*

These distributions show no excess around 650 GeV, from which one would naïvely conclude that the excess measured in that region does not come from a scalar meson. This is indeed only indicative given that the discriminant cut only keeps a small fraction of the selected samples. A tensor decaying into ZZ has a forward peaked distribution, $d\sigma/dcos\theta_Z \sim (3cos^2\theta_Z - 1)^2$, in the ultra-relativistic case (figure 9), which clearly shows this peaking, in similarity to the DY background. It is therefore tempting to believe that the discriminating selection applied in [32] would cause damages if the excess around 650 GeV originates from a tensor particle.

The following picture, extracted from [31], summarizes a rather complex situation for a tensor scenario.

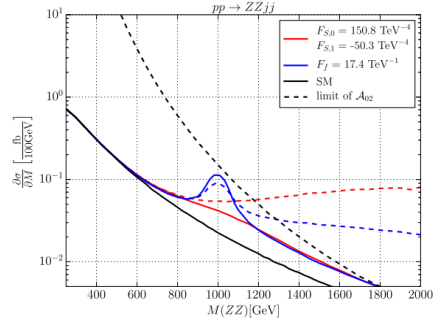

*Figure 10: Differential cross section for an isoscalar tensor resonance with mass 1000 GeV and width 100 GeV, for VBF->ZZ. Solid line in blue : unitarized results, dashed line uncorrected results.*

From this figure, one can retain that the unitarized solution, in blue, resembles to a genuine BW without blowing up at high energy.

Retaining only the resonant Born term one has the following amplitude:

A(w+w- -> w+w-)= -(1/24)$F_f^2$ $s^2 P_2(s,t,u)/(s^2-m^2+im\Gamma_{tot})$ with $P_2=[3(t^2+u^2)-2s^2]/s^2$

Assuming WW, ZZ and hh contributions, the total width reads:

$$\Gamma_{tot}=m^3 F_f^2/\pi 960$$

The coupling constant $F_f$ is adjustable, such that knowing m=650 GeV one can adjust it to reproduce the observed value $\Gamma_{tot}$=100 GeV. This gives a total VBF cross section of ~60 fb for T(650), below observation.

Note however that one should also add the top quark coupling which provides a ggF contribution. Assuming $\Gamma_{tt}$=15 GeV, one gets VBF+ggF~500 fb, in better agreement with observations.

In the tensor case, one has ZZ/WW=0.5 which is compatible with observations as noted in section VII.

Also, a tensor particle naturally couples to h(125)h(125), h(95)h(95) and h(125)h(95), the later not being affected by a reduction of 2 for identical bosons, which may explain the CMS first observation of H(650)->h(125)h(95).

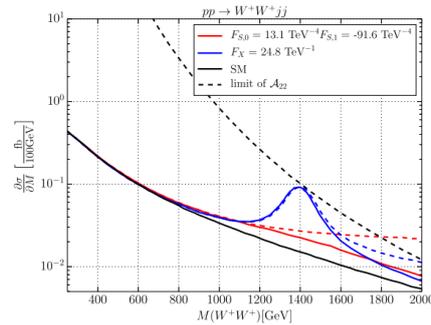

*Figure 11: Same as 9 in the case of an isotensor resonance with m=1400 GeV and a width of 140 GeV.*

What happens in the **isotensor** case? Similar features to figure 10 are observed in figure 11. In this case the total width is given by:

$$\Gamma_{tot}=m^3 F_X^2/\pi 3840$$

The total VBF cross is ~60 fb, again requiring a large ggF contribution to reproduce de data.

One usually assumes that T, T+ and T++ have the same mass, but this hypothesis may be relaxed as is the case in the GM model.

*A Kaluza-Klein scenario ?*

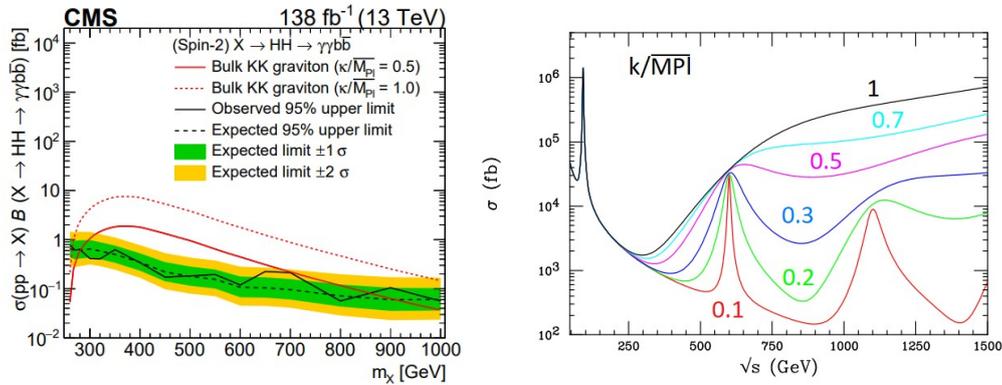

*Figure 12. Left, from CMS [34]: expected and observed 95% CL upper limit on resonant production cross section for spin-2 pp->X->HH->γγbb signal hypothesis. The red lines show the bulk KK graviton predictions for two values of k/$\overline{MPl}$ . Right, from [35]: cross section for e+e- -> G(600) -> µ+µ- versus k/$\overline{MPl}$.*

Leaving generalities, one may ask if such a tensor could fit into some specific BSM models. What comes to mind are models with extra-dimensions which predict heavy **Kaluza Klein excitations of gravitons**. Recently [34] CMS has given an interpretation of the excess observed in γγbb at 650 GeV in terms of a **bulk KK graviton** with k/$\overline{MPl}$=0.5 as shown in figure 12. This value is on the high side of what the theory expects as pointed out in [35]. The plot from [35] is only shown for illustration since one does not expect the KK bulk graviton to significantly couple to light fermions. Several KK graviton recurrences could be present with a hierarchy given by: $x^G_i$=3.83, 7.02, 10.17, 13.32.

One could speculate that T(320) and T(650) are the **two first recurrences** given that for mG2=650 GeV, one predicts mG1=355 GeV. One also predicts mG3=950 GeV, clearly outside the GM mass domain.

**h(95)** could then be a **radion,** a scalar expected in an extra dimension scenario [40]. [41] claims that KK gravitons naturally couple to scalars.

 **T(320)->h(125)h(95)** and **h(95)h(95)** should also be searched for, although one may speculate that, given the poor bb mass resolution, these modes are already included in the bbbb analysis of ATLAS.

The last word pertains to the **angular distribution** of X(650)->ZZ, markedly different for J=0 and J=2.

## A5. Extraction of the GM parameters

This appendix intends to provide an evaluation of the GM parameters from the various observables. The aim is to predict the BR of H5++(450) into H3+(130)H+(130) and of H5+(130) into H3+(130)A3(151). To do this, we need to extract the parameters λi and M1 and M2.

From the SR alone we are able to extract the vacuum expectation u of the two triplets and the BR into VV and VH:

| Channel | u GeV | $s_H$ | BR(VV) % | BR(VH) % |
|---|---|---|---|---|
| H++ | 70±12 | 0.80±0.1 | 9 | 12.5 |
| H+ | 80±13 | 0.90±0.2 | 15 | 17 |

The total widths of these resonances deduced from the SR, ΓH++~160 GeV and ΓH+~80 GeV, imply that the width of H5++ into H3+H3+ should be or order 100 GeV. The couplings [16] of these modes are given by:

$$C_{H_3^0 H_3^+ H_5^{+*}} = -i\frac{\sqrt{2}}{v^2}\left[2(\lambda_3 - 2\lambda_5)v_\phi^2 v_\chi - 8\lambda_5 v_\chi^3 + 4M_1 v_\chi^2 + 3M_2 v_\phi^2\right],$$

$$C_{H_3^+ H_3^+ H_5^{++*}} = -\frac{2}{v^2}\left[2(\lambda_3 - 2\lambda_5)v_\phi^2 v_\chi - 8\lambda_5 v_\chi^3 + 4M_1 v_\chi^2 + 3M_2 v_\phi^2\right],$$

Unitarity constraints prevent large values of the λi, with the notable exception of **λ5** which can be [16] as large as ±4π. M1 and M2 can also be large, subject to the constraints:

$$m_5^2 = \frac{M_1}{4v_\chi}v_\phi^2 + 12 M_2 v_\chi + \frac{3}{2}\lambda_5 v_\phi^2 + 8\lambda_3 v_\chi^2,$$

$$m_3^2 = \frac{M_1}{4v_\chi}(v_\phi^2 + 8v_\chi^2) + \frac{\lambda_5}{2}(v_\phi^2 + 8v_\chi^2) = \left(\frac{M_1}{4v_\chi} + \frac{\lambda_5}{2}\right)v^2.$$

where **m5~450 GeV** and **m3~140 GeV**. The vacuum expectations are trivially related to u and sH:

**vϕ=vcH=148 GeV, vχ=u=70 GeV**

Knowing the properties of h(95) and h(125), one can easily derive λ1=0.07 and λ2=−1.4 and the combination λ3+3λ4=2.6. From this equation and the unitarity and 'bound from below' constraints [16], one can define an allowed region around λ3~−0.94 , λ4~1.03, as shown in figure 13.

An extra condition is needed to extract the three missing quantities: M1, M2 and λ5. We simply impose that the width of H5++ into H3+H3+ equals the missing width.

We find a solution, which gives Γ$_{H++\to H3+H3+}$ =100 GeV and Γ$_{H+\to A3H3+}$ = 50 GeV in agreement with the SR. Note that the H+ width into A3H3+ is reduced with respect to H3+H3+ due to a **phase space** suppression of H+(375)→A3(151)H3+(130) with respect to H++(450)→ H3+(130)H3+(130).

| u GeV | λ1 | λ2 | λ3 | λ4 | λ5 | M1 GeV | M2 GeV |
|---|---|---|---|---|---|---|---|
| 70 | 0.07 | -1.4 | -1.06 | 1.25 | -6.3 | 950 | 400 |

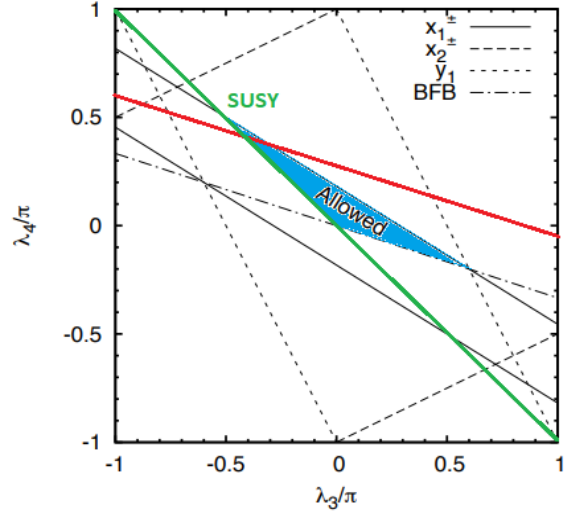

*Figure 13: The blue zone combines unitarity and 'bound from blow' constraints [16] and the red line the relation λ3+3λ4=2.69.*

Admittedly, the large value of **λ5** calls for a comment. While it is compatible with unitarity constraints, its large value main imply that HO corrections could be important.

Note that the table suggests that the relation λ3~– λ4 predicted within the **SUSY** version of GM proposed in [49] is verified (green line in figure 13), while the two others: λ1~3/4λ2 and λ5~–4λ2+2√λ2λ4, are violated.

Again one should not forget that our results rely on the hypothesis of a minimal GM content, which is not the case for a SUSY extension.

### *h95 →2γ , h125->Zγ and κλ*

From above quantities, one can determine the widths into two photons of the two lightest scalars normalized to the SM. To do this one needs to extend the GM formalism with the addition of an extra doublet which is a task for future developments. For the time being, we keep the standard GM formalism, choosing sinα=0.81 and cosα=-0.58 to insure that **μ125γγ~1.** One then finds:

**μ95γγ=0.22** as compared to **0.24±0.09** from [48]

in surprisingly good agreement given the large value of λ**5,** indicating that HO corrections are small. This agreement is non trivial since, as already pointed out, in absence of the charged Higgs contributions, one finds that μ95γγ~1, suggesting the relevance of these charged scalar contributions.

For what concerns **μ125γγ**, there is a cancellation between the contribution of H3+(130) and that of H5+(375) and H5++(450). For h125→Zγ, one finds:

**μ125Zγ=2.3** as compared to **μ125Zγ=2.2±0.7** from [45]

again in good agreement with, admittedly, a large experimental error.
This measurement shows the strongest predicted deviation with respect to the SM.

One can also compute the triple gauge coupling for h125, predicting:

$$\kappa\lambda = \lambda hhh/SM = 5$$

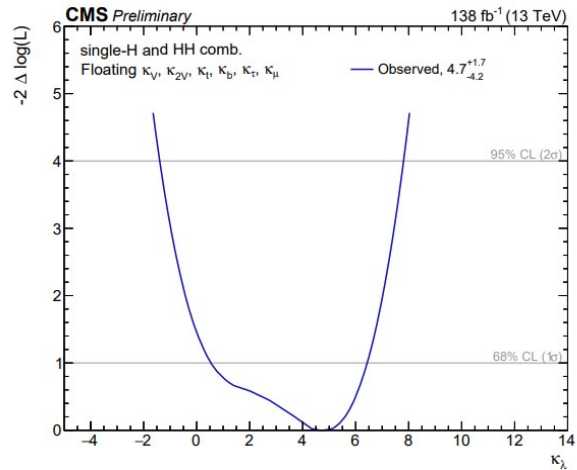

*Figure 14: A recent compilation from CMS [50] on the self-coupling parameter κλ normalised to the SM. Observed likelihood scans of κλ assuming κV, κ2V, κt , κb , κτ , and κμ as unconstrained nuisance parameters*

Figure 14 shows the present status of this measurement. It does not contradict our prediction. If true, this GM prediction should be confirmed with HL-LHC.

Again, these results are obtained with an **approximate treatment of mixing effects** for h(125) and h(95), meaning that above agreements could be purely accidental.